\documentclass[conference]{IEEEtran}
\usepackage{amssymb}
\usepackage{paralist}
\usepackage{url}
\usepackage{color}

\usepackage{graphicx}
\newcommand{\ndnname}[1]{{\small \tt #1}}
\newcommand{\descr}[1]{\vspace{0.22cm} \noindent\textbf{#1}}

\title{DoS \& DDoS in Named-Data Networking}
\date{}
\author{Paolo Gasti \and Gene Tsudik \and Ersin Uzun \and Lixia Zhang}
\begin{document}
\maketitle

\begin{abstract}

With the growing realization that current Internet protocols are
reaching the limits of their senescence,  a number of on-going
research efforts aim to design potential next-generation Internet
architectures. Although they vary in maturity and scope, in order to
avoid past pitfalls, these efforts seek to treat security and privacy
as fundamental requirements. Resilience to Denial-of-Service (DoS)
attacks that plague today's Internet is a major issue for any new
architecture and deserves full attention.

In this paper, we focus on DoS in a specific candidate next-generation 
Internet architecture called {\em Named-Data
Networking} (NDN) -- an instantiation of Information-Centric Networking approach. By stressing content
dissemination, NDN appears to be attractive and viable approach to
many types of current and emerging communication models. It also
incorporates some basic security features that mitigate certain
attacks. However, NDN's resilience to DoS attacks has not been
analyzed to-date. This paper represents the first step towards
assessment and possible mitigation of DoS in NDN. After identifying
and analyzing several new types of attacks, it
investigates their variations, effects and counter-measures.
This paper also sheds some light on the long-standing debate
about relative virtues of self-certifying, as opposed to human-readable,
names.

\end{abstract}


\section{Introduction \label{intro}}

The Internet clearly represents an overwhelming and unique global success story.
Billions of people worldwide use it to perform a wide range of
everyday tasks. It hosts a large number of information-intensive
services and interconnects many millions of wired, wireless,
fixed and mobile computing devices.
The Internet also serves as a means of disseminating
enormous (and ever-increasing) amounts of digital content. Since
its inception, the amount of data exchanged over the Internet has
witnessed exponential growth. Recently, this growth intensified due
to increases in: (1) distribution of multimedia content, (2)
popularity of social networks and (3) amount of user-generated
content. Unfortunately, the same usage model that fostered Internet's
success is also exposing its limitations. Core ideas of today's
Internet were developed in the 1970-s, when telephony -- exemplified by a
point-to-point conversation between two entities -- was the only
successful example of effective global communication
technology. Moreover, original Internet applications were few and
modest in terms of bandwidth and throughput requirements, e.g.,
store-and-forward email and remote computer access.

The way people access and utilize the Internet has changed
dramatically since the 1970-s and today the Internet has to
continuously accommodate new services and applications as well as
different usage models. To keep pace with changes and move the
Internet into the future, a number of research efforts to design new
Internet architectures have been initiated in recent years.

Named-Data Networking (NDN) \cite{NDN} is one such effort. NDN is an
on-going research project that aims to develop a candidate
next-generation Internet architecture. It instantiates the so-called
Content-Centric (CCN) or Information-Centric (ICN) approach
\cite{gritter2001architecture,Jacobson2009,koponen2007data} to
networking. NDN explicitly names content instead of physical locations
(i.e., hosts or network interfaces) and thus transforms content into
a first-class entity. NDN also stipulates that each piece of named
content must be digitally signed by its producer. This allows
decoupling of trust in content from trust in the entity that might
store and/or disseminate that content. These NDN features facilitate
automatic caching of content to optimize bandwidth use and enable
effective simultaneous utilization of multiple network interfaces.

NDN has been demonstrated as a viable and attractive architecture for content
distribution \cite{Jacobson2009} as well as real-time
\cite{VanSmBriPlStThBra09-Voice} and anonymous communication
\cite{andana}. 
A number of other NDN-related research efforts are also underway.

\subsection{DoS and DDoS}
In recent years, denial of service (DoS) and distributed denial of
service (DDoS) attacks have become more and more common and notorious. In the
latter, the adversary exploits a large number of compromised hosts
(zombies), that surgically aim their attacks at specific target(s).
Although DDoS attacks are generally easy to instantiate and require
little technical sophistication on the part of the adversary, they
are often very effective and difficult to mitigate. (NOTE: hereafter, we
use the term {\em DoS} to include both single-source and multi-source
-- i.e., DDoS -- denial-of-service attacks.)

We believe that any new Internet architecture should: (1) be
resilient to existing DoS attacks, or at least limit their
effectiveness, (2) anticipate new attacks that take advantage of its
idiosyncrasies, and (3) incorporate basic defenses in its design.

To the best of our knowledge, there has been no scientific and
systematic assessment of how NDN fares with respect to DoS attacks.
We believe that such assessment is both timely and very important. While NDN
appears to be quite efficient in terms of content distribution between
well-behaved (honest) entities, it is unclear how it would cope with
malicious parties. This paper tries to address these issues by
analyzing the impact of current DoS attacks on NDN, identifying new
attacks that rely on NDN features, and proposing some
countermeasures. We emphasize that this paper should be seen not as a
comprehensive treatment of security (or even DoS) in NDN. Instead, it 
represents a first step towards identifying DoS attacks as well as their impact
and securing NDN against them.

\subsection{Anticipated Contributions}
The main goal of this paper is to explore, and evaluate
effects of DoS attacks against NDN. We believe that only by better understanding the
effects and repercussions of these attacks we can begin to develop
sensible and effective countermeasures. Anticipated contributions are as follows:
\begin{compactitem}
\item We show that DoS attacks effective in the current IP-based Internet
    are largely ineffective against NDN. This is not really
    surprising since current attacks are very much tailored to
    TCP/IP,  current routing protocols, HTTP, DNS, etc.
\item We identify and describe two new and major types of NDN-specific
    DoS attacks, based on: (1) {\bf interest flooding}, and (2)
    {\bf content/cache poisoning}. We argue that both types of attack require 
    long-term exploration.
\item We then analyze the impact of several flavors of both attack types and
    propose a set of potential counter-measures.
\end{compactitem}
\medskip
\descr{Organization:} The rest of the paper is structured as follows: Section~\ref{sec:ndn} provides an 
overview of NDN. We discuss how NDN copes with existing DoS/DDoS attacks in 
Section~\ref{sec:current}, and examine new -- NDN-specific -- ones in Sections \ref{sec:flooding} and 
\ref{sec:poisoning}. Section~\ref{sec:relw} overviews related work. Finally, we conclude in Section 
\ref{sec:conclusions}

\section{NDN Overview}
\label{sec:ndn}
As mentioned in Section \ref{intro}, 
NDN \cite{NDN,Jacobson2009} is a network architecture based on named content.
Rather than addressing content by its location, NDN refers to it by
name. A content name is composed of one or more variable-length
components that are opaque to the network. Component boundaries are
explicitly delimited by ``\ndnname{/}'' in  the usual representation.
For example, the name of a CNN news content for May 20, 2012 might
look like: \ndnname{/ndn/cnn/news/2012may20}. Large pieces of content
can be split into fragments with predictable names: fragment $137$ of
Alice's YouTube video could be named:
\ndnname{/ndn/youtube/alice/video-749.avi/137}. Since NDN's main
abstraction is content, there is no explicit notion of ``hosts'',
albeit, their existence is assumed.

Communication adheres to the {\em pull} model: content is delivered
to consumers only upon explicit request. A consumer requests content
by sending an {\em interest} packet. If an entity (a router or a host)
can ``satisfy'' a given interest, it returns the corresponding {\em data packet}.
Interest and content are the only types of packets in NDN. A
content packet with name X is  never forwarded or routed
unless it is preceded by an interest for name X. (Strictly speaking,
content named X$'\neq$ X can be delivered in response to an interest
for X, but only if X is a prefix of X$'$.)

NDN routers must include of the following components:
\begin{compactitem}
\item {\em Content Store} (CS), 
used for content caching and retrieval; 
\item {\em Forwarding Interest Base} (FIB), that
contains a table of name prefixes and corresponding outgoing interfaces 
(to route interests);
\item {\em Pending Interest Table} (PIT) -- a table containing a set of currently
unsatisfied  interests and their corresponding incoming interfaces; 
\end{compactitem}
When a router receives an interest for name X and there are no
pending interests for the same name in its PIT, it forwards the
interest to the next hop, according to its FIB. For each forwarded
interest, a router stores some amount of state information, including
the name in the interest and the interface on which it arrived.
However, if an interest for X arrives while there is already an entry
for the same name in the PIT, the router collapses the present
interest (and any subsequent ones for X) storing only the interface
on which it was received. When content is returned, the router
forwards it out on all interfaces from which an interest for X has
arrived and flushes the corresponding PIT entry. Since no additional
information is needed to deliver content, an interest does not carry
a ``source address''.

Any NDN router can provide content caching through its CS. The
size of CS is limited only by resource availability.
Consequently, content might be fetched from any number of network
caches, rather than from its original producer. Hence, NDN has no
notion of ``destination addresses''. 

NDN deals with content authenticity and integrity by making digital
signatures mandatory for all content. A signature binds content with
its name, and provides origin authentication no matter how, when or
from where it is retrieved. Public keys are treated as regular
content: since all content is signed, each public key content is
effectively a simple ``certificate''. NDN does not mandate any
particular certification infrastructure, relegating trust management
to individual applications.

Content objects are named data packets.\footnote{In the rest of the paper, we use the terms content object and data packet interchangeably.} 
Fields of a data packet include~\cite{content-objects}:
\begin{compactitem}
\item {\tt Signature}: public key signature (e.g., RSA or DSA) 
computed over the entire data packet, including its name.
\item {\tt Keylocator}: references the key 
needed to  verify the content signature. 
This field can contain one of the following: (1) verification (public) key; 
(2) certificate containing verification key; or (3) NDN name referencing verification key. 
\item {\tt PublisherPublicKeyDigest}: hash of the data packet producer's public key.
\end{compactitem}
In addition to the name of requested content, an interest packet carries 
several fields~\cite{interests}. In this paper, we are interested in the following 
(others are omitted for clarity):
\begin{compactitem}
\item {\tt PublisherPublicKeyDigest}: this optional field contains the 
hash of the producer's public key for the requested piece of data.
\item {\tt Exclude}: an optional field that embodies a description of 
name components that should not appear in the data packet in response to the interest.
\item {\tt AnswerOriginKind}: encodes determines whether the answer 
to an interest can retrieved from a CS or must be generated by the producer.  
\item {\tt Scope}: limits where the Interest may propagate; Scope 0 and 1 limit 
propagation to the originating host; Scope 2 limits propagation to no further than the next host.
\end{compactitem}

\section{NDN and Current DoS Attacks\label{sec:current}}
We now consider how NDN fares against DoS attacks. We initially 
focus on current DoS types often encountered on the Internet. 
Next, we analyze the impact of these attacks on NDN and discuss our 
findings. Then, in the next two sections, we identify some new DoS attacks that take advantage 
of NDN features, and discuss tentative countermeasures.

There is a wide variety of DoS attacks targeting different network and host 
resources, protocol layers as well as specific software. A typical attack is 
composed of three elements:
(1) a set of zombies under control of a master node or nodes,
(2) master node(s) controlling the zombies, and (3) a set of victim hosts and/or routers.

A common way to obtain zombies to remotely exploit software vulnerabilities 
and inject malware into a set of unpatched hosts. NDN makes this somewhat harder 
-- yet far from impossible -- because hosts are not directly addressable. 
However, there is no indication that software vulnerabilities will 
disappear or even decrease significantly over time. Thus,
we assume that adversaries will continue to have access to large
numbers of zombies by means such as exploitation of un-patched vulnerabilities 
or phishing.

\subsection{Impact of Current Attacks on NDN}
We now examine some popular types of DoS attacks that work 
against current TCP/IP-based Internet and assess their putative effects on NDN.

\descr{Reflection Attacks. }
A reflection attack involves three parties: the adversary, a victim host, and a 
set of secondary victims (reflectors). The goal of the adversary is to use the reflectors to overwhelm the 
victim host with traffic.
To do so, a reflection attack uses IP packets with forged addresses: the adversary replaces its own source 
address with the address of its
intended victim, and sends these packets to the secondary victims. 
Responses to such packets are not routed back to the adversary, and overwhelm the victim instead. 
To be effective, such attacks require some form of
amplification, i.e., the amount of data used by the adversary to
perform the attack must be significantly smaller than the amount of
data received by the victim.

NDN is generally resilient to this type of attack due to the
symmetric nature of the path taken by each interest and the
corresponding content. A content packet must follow, in reverse, the
path established by the preceding interest. However, note that an NDN
router is allowed to broadcast an incoming interest on {\bf some or
all} of its interfaces. (In other words, an interest broadcast can
occur at any hop). Nonetheless, even if each hop (including the
consumer) broadcasts the interest, the maximum number of content
copies a consumer can receive is bounded by the number of its interfaces,
and not by the number of entities that receive an interest.
Consequently, the only effective reflection-style attack requires the
adversary to be on the same physical network (e.g., Ethernet or WLAN)
as the intended victim. Then, the adversary broadcasts one or more
interests on all available interfaces (with the victim's 
layer-2 address as the source) and the victim subsequently receives multiple
copies of requested content. This attack seems somewhat effective
since the adversary's amplification factor is based on the
comparatively small size of interests with respect to that of
content. However, NDN routers incorporate a useful suppression feature:
\begin{quote} {\em
Whenever an NDN router ``overhears'' a content packet on a broadcast
interface $IF_x$ for which it has a current PIT entry with the
incoming interface $IF_x$, it caches the said content and flushes the
PIT entry. Note that another copy of the same content might be later
delivered via the outgoing interface ($IF_y$) for that PIT entry; it
will be discarded since, by that time, the corresponding PIT entry
will have been flushed. \rm}
\end{quote}
This feature ensures that multiple NDN routers on the same broadcast
domain do not serve the same content more than once, even if the
original interest was broadcast.

\descr{Bandwidth Depletion. }
In a typical coordinated distributed attack, adversary-controlled
zombies flood their victims with IP traffic in order to saturate their network resources. The usual goal is to make the victims
unreachable by others and/or, more generally, to inhibit victims'
ability to communicate. Normally, such attacks are carried out via
TCP, UDP or ICMP and rely on sending a stream of packets to the
victim at the maximum data rate.\footnote{TCP-based attacks also exploit
connection-based nature of the protocol: each packet sent by zombies
tries to open a new connection, which, in turn, requires the victim
to create and store corresponding state, thus saturating its
resources.}

A similar kind of attack can be mounted against NDN by directing a
large number of zombies to request existing content from a certain victim.
However, it is easy to see that the effectiveness of this attack
would be very limited. Once the content is initially pulled from its
producer, it is cached at intervening routers and  subsequent
interests retrieve it from these routers' caches. Therefore, the network itself 
would limit the number of interests that reach the victim.

\descr{Black-Holing and Prefix Hijacking. }
In a prefix hijacking~\cite{hijacking} attack, a misconfigured,
compromised or malicious autonomous system (AS) advertises invalid
routers so as to motivate other AS-s to forward their traffic to it.
This can result in so-called ``black-holing'' whereby all traffic sent
to the malicious AS is simply discarded. This attack is effective in
IP networks, since, once routing information is polluted, it is difficult for 
routers to detect, and recover from, the problem.
While countermeasures have been proposed (e.g.,
\cite{lad-PHAS}) this remains a serious threat to the current Internet.

NDN is resilient to black-holing implemented via prefix hijacking.
NDN routers have access to strictly more information than their IP
counterparts and can use such information to detect anomalies in the
content distribution process. Since each content follows the
same path as the interest that requested it, the number 
of unsatisfied (expired) interests can be used to determine whether a
particular prefix has been hijacked. Also, NDN routers maintain
statistics about performance of each link and interface with respect
to a particular prefix and change their forwarding strategy according
to such statistics. 

Loop detection and elimination allows routers to explore topological
redundancy through multipath forwarding. Multipath routing further
reduces impact of prefix hijacking, since it allows routers to try
alternative paths as a reaction to attacks. This increases the
probability of forwarding interests through a path that has not been
affected by the attack. In contrast with IP, advertising fake routes in NDN does
not allow the adversary to mount a loop-holing attack

To sum up, the NDN forwarding plane allows routers to: (1) quickly
(i.e., at RTT scale) detect and react to network failures; (2)
detect and circumvent hijacking attacks and (3) incorporate
congestion into forwarding decisions.

\descr{DNS Cache Poisoning. }
In the current Internet, DNS servers translates human-readable names
to the corresponding IP address and vice-versa. For performance
reasons, DNS servers usually store the output of previous requests in
their cache. There is a well know attack, called DNS cache
poisoning~\cite{dns-poisoning}, which allows the adversary to insert
corrupted entries in a DNS server's cache in order to control the
server responses for a set of DNS names. The best countermeasure
against this attack is the use of the DNS Security Extensions
protocol, i.e., DNSSEC~\cite{dnssec}; however, as of today DNSSEC has
not been widely deployed on the Internet.

Packet names in NDN are routed directly, rather than being converted
to addresses. While this implies that there is no need for services
that perform name resolution (and therefore such service can not be
corrupted), it still is possible to conceive an attack analogous to
DNS poisoning on NDN. We believe that the closest counterpart of DNS
cache poisoning in NDN is a combination of route hijacking and
content poisoning: the adversary would force a routing change (if
necessary) that allows it to be on the path for a set of namespaces
that are going to be affected by the attack. Then, it answers
interests with data packets carrying an arbitrary payload.

Intuitively, signatures on data packets allow consumers to
determine whether the content has been poisoned. Sections
\ref{sec:poisoning} and following are devoted to this kind of attack
and its countermeasures.

\subsection{Towards New DoS Attacks}
Despite their seeming lack of effectiveness or greatly reduced
impact, variations of aforementioned current DoS attacks might be quite
effective against NDN. 

Recall that two key features distinguish current Internet 
routers from their NDN counterparts: (1) pending interest state (PIT entries) needed to 
perform content routing and (2) the use of content caches.
In subsequent sections, we describe two classes of attacks -- Interest Flooding 
and Content/Cache Poisoning -- that capitalize on these NDN features.
In addition, we anticipate other types of DoS-related NDN attacks, 
such as those that focus on routing. However, we believe that attacks 
targeting these two specific key features of NDN represent most serious
and immediate threats, thus warranting an in-depth investigation.

\section{Interest Flooding\label{sec:flooding}}
Routing of content is performed using state in PIT-s established by
interests, i.e., the name of each incoming content packet is used
to look up the PIT and identify a corresponding entry.  
The adversary can take advantage of this state to mount an
effective DoS attack, which we term ``interest flooding''. In such an attack,
the adversary (controlling a set of possibly geographically distributed zombies)
generates a large number of closely-spaced interests, aiming to (a)
overwhelm (PIT-s) in routers, in order to 
prevent them from handling legitimate interests, and/or (b)
swamp the targeted content producer(s). 
Since NDN interests lack source address and are not secured (e.g., not signed) 
by design, it is difficult to determine the attack originator(s)
and take targeted countermeasures.

We identify three types of interest flooding attacks, based on the
type of content requested: (1) existing or static, (2)
dynamically-generated, and (3) non-existent (i.e., unsatisfiable interests). 
In all cases, the adversary uses zombies to generate a large number
of interests requesting content from targeted producers.
While attacks (1) and (3) are mostly aimed at the network infrastructure,
(2) affects both network and application-layer functionalities.

Similarly to bandwidth depletion attacks discussed in the previous section,
the impact of type (1) attacks is quite limited since
in-network content caching provides a built-in countermeasure.
Suppose that there are several zombies, each with independent path to
the targeted producer. After the initial ``wave'' of  interests from
these zombies, content settles in all intervening routers' caches.
Subsequent interests for the same content do not propagate to the
producer(s) since they are satisfied via cached copies.

In case of type (2) attacks, benefits of in-network content caching are
lost. Since requested content is dynamic, all interests are routed to
content producer(s), thus consuming bandwidth and router PIT state. Also,
if generating dynamic content is expensive -- signing content is
a good example relatively expensive per-packet operation -- content producers might waste
significant computational resources. (One concrete example of this
attack class could target a web server that allows site-wide
searches: each zombie issues an interest that requests the victim
server to search its entire site for a random string.)
The direct outcome of this attack type is that the producer wastes
resources to satisfy malicious, rather than legitimate, interests. 
The impact on routers varies with their distance from the
targeted content producer: the closer a router is to the producer,
the greater the effect on its PIT due to more concentrated attack traffic.

Type (3) attacks involve zombies issuing unsatisfiable interests for
non-existent content. Such interests can not be collapsed by routers, 
and are routed to the targeted content producers. 
The latter can quickly ignore such interests
without incurring significant overhead. However, such
interests will linger and take up space in router
PIT-s until they eventually expire. We consider routers to be primary
intended victims of this attack type. Given an existing prefix \ndnname{/ndn/prefix}, 
there are several easy ways to construct unsatisfiable interests: 
\begin{compactenum}
\item Set the name in the interest to: \ndnname{/ndn/prefix/nonce}, 
where the suffix \ndnname{nonce} is a random value.  Since NDN performs 
longest-prefix matching, interests will be forwarded all the way to the producer
and never satisfied.
\item Set the {\tt PublisherPublicKeyDigest} field to a random value. 
Since no public key would match this value, the interest will remain unsatisfied.
\item Set the interest {\tt Exclude} filter to exclude all existing content starting with 
\ndnname{/ndn/prefix}, e.g., by using a Bloom Filter with almost all bits set to 1. 
Such an interest can not be satisfied since it simultaneously requests {\em and} 
excludes the same content.
\end{compactenum}
In fact, there is another way of mounting type (3) attacks that does not even 
require the adversary to recruit any zombies. 
Recall that each data packet has a {\tt KeyLocator} field (see Section~\ref{sec:ndn}). 
This field can contain the NDN name of the data packet's verification key. This feature
can be abused to mount a type (3) attack as follows: 
\begin{quote} {\em
The adversary publishes a large amount of content (e.g., a video), that is split into 
numerous content packets. Each packet references a distinct non-existent verification key, 
such that an interest for this key is not satisfiable. }
\end{quote}
To increase the effectiveness of this attack, all spurious key names must be routed towards the same 
producer. This way, a large number of unsuspecting consumers are ``forced'' to issue interests 
that overwhelm one or more victim's PIT-s. We point out that NDN automatically 
mitigates such attacks through interest collapsing. Since the pool of fake key-names is limited by the 
number of packets comprising malicious content, rather than by the combined bandwidth of the 
zombies, interests with the same name will be collapsed by routers. Therefore, the total number of 
fake key-names represents an upper bound on the amount of space that can be consumed in 
the victims' PIT-s.

\subsection{Tentative Countermeasures}
\label{sec:flood-countermeasures}
We consider several potential countermeasures. The ease of 
interest flooding attacks is partly due to lack of
authentication of interests. Anyone can generate a stream of
interests and a given NDN router only knows that a particular
interest entered on a specific interface. There is no other
information about its source. One trivial solution might be to require
signatures on interests. However, this would immediately
raise serious privacy concerns, as discussed in
\cite{andana} and would also introduce new DoS vulnerabilities due to
the computational overhead of signature verification.

On the other hand, we believe that potential problems and DoS attacks
due to interest flooding can be addressed without requiring source
authentication. This is because NDN routers are stateful
and can learn much more information about carried traffic than 
their current IP counterparts. 

\descr{Router Statistics.}
NDN routers 
can easily keep track of unsatisfied (expired) interests and use this
information to limit the following:
\begin{compactitem}
\item{\# of pending interests per outgoing interface:} NDN creates flow
    balance between interests and content. For each interest sent
    upstream, at most one data packet satisfying that interest
    can flow downstream. Based on that property, it is trivial
    for each router to calculate the maximum number of pending
    interest per outgoing interface that the downstream
    connection can satisfy before they time out. Thus, a router
    should calculate and never send more interests than an
    interface can satisfy based on average content package size,
    timeout for interests and bandwidth-delay product for the
    corresponding link.
\item{\# of interests per incoming interface:} From the same flow
    balance principle, a router can easily detect when a
    downstream router is sending too many interests that can not
    be all satisfied due to the physical limitations of the
    downstream link.
\item{\# of pending interests per namespace:} When a certain prefix is
    under a DoS attack, routers on the way (especially those
    closer to the data producer) can easily detect unusual number
    of unsatisfied interests in their PIT-s for that prefix.
    In that case, routers can limit the total number of pending
    interests for that prefix and throttle down the number of
    pending interests for incoming interfaces that have sent too
    many unsatisfied interests for that prefix.
\end{compactitem}
Although these countermeasures seem intuitive and possibly effective, we believe 
that implementing and testing them will be quite difficult. Most of all,
combining the above three limiting strategies into one algorithm and choosing
appropriate parameters for maximum effectiveness against attacks and
minimum impairment of legitimate traffic is a challenge.
We leave the design and testing of the actual algorithms to future work. 

\descr{Push-back Mechanisms.}
We also consider one router-based countermeasure to interest flooding -- 
a push-back mechanism that allows routers to isolate attack source(s).
When a router suspects an on-going attack for a particular namespace
(e.g., when it reaches its PIT {\em quota} for that namespace on a given
interface), it throttles any new interests for that namespace and reports
this to routers connected on that interface. These routers, in turn, can
propagate such information upstream towards offending interfaces, 
while also limiting the rate of forwarded interests for the namespace
under attack. The goal is to push an attack back all the way to 
its source(s), or at least to the location where it is detectable.
This countermeasure can be implemented without any modifications to the current 
NDN infrastructure.

\section{Content/Cache Poisoning}
\label{sec:poisoning} 
We now shift focus to DoS attacks that target content.
In this context, the adversary's goal is to cause routers to forward and cache
corrupted or fake data packets, consequently preventing consumers from 
retrieving legitimate content. We say that  a data packet is {\rm corrupted} if
its signature is invalid. Whereas, a data packet is {\em fake} if it has a valid
signature, however, generated with a wrong (private) key.

As mentioned in Section \ref{sec:ndn}, all data packets in NDN are signed. This
provides the following security guarantees: 
\begin{compactitem}
\item {\em Integrity} -- a valid signature guarantees that the signed data packet is intact; 
\item {\em Origin Authentication} -- since a signature is uniquely bound to 
the public key of the signer, anyone can verify whether content originates with the claimed producer;
\item {\em Correctness} -- a signature binds data packet name with its
payload, thus allowing a consumer to securely determine whether a 
data packet  is a ``correct answer'' for the interest that requested it.
\end{compactitem}
Consumers are expected to perform signature verification on every data packet
before accepting it.  Also, any NDN router can elect to perform signature verification
for any content it forwards and caches. 
Upon receiving and identifying a corrupted or fake data packet, a consumer can 
re-request a different (possibly valid) copy of the same data packet using the {\tt Exclude} 
field in NDN interest packet.
 
In theory, content signatures provide an effective and simple means for detecting 
content poisoning attacks, since ``bad'' content can be easily identified via signature
verification. In other words, NDN {\em should be} immune to content poisoning attacks.
However, in practice, this assertion might not hold. While a consumer can afford
to verify all content signatures, NDN routers face two challenges: (1) {\bf signature 
verification overhead}; and (2) {\bf trust management}, i.e., what key should be 
used to verify a given data packet?

While routers can choose to verify signatures on each data packet they 
forward and/or store, for performance reasons, they are not {\em required} 
to do so. Our tests show 
that an optimized software implementation of RSA-1024 signature verification running 
on Intel Core 2 Duo 2.53 GHz CPU allows us to verify about 150 Mbps of 
traffic, assuming $1,500$-byte content packets. (Smaller packets would impose 
even higher verification overhead). 
Note that we use the smallest possible RSA public exponent -- $3$ -- thereby incurring 
only two modular multiplications per signature verification. Routers with multiple 
Gigabit-speed (or faster) interfaces would need an unrealistic amount of 
computing power to verify packets at wire rate. 

Content signatures also trigger the issue of global trust management architecture.
Without it, routers can not determine the public key needed to verify the data packet 
signature. This creates a tension between flexibility (since an application can adopt
an arbitrary trust model for its content) and security
(any NDN router must be able to, if its chooses, verify any data packet's signature).
Even though each NDN data packet contains a reference to its signature
verification (public) key, such references can not be trusted as they can be easily
abused by the adversary.

\subsection{Attack Variants}
\label{sec:attack-vectors}
The impracticality of NDN routers verifying all signatures on forwarded or cached 
data packets opens the door for content poisoning
attacks. As mentioned before, one can not push poisoned content 
unilaterally, i.e., without any prior interest requesting that 
content. Consequently, we identify two attacks variants:
\begin{compactenum}
\item Suppose that the adversary {\em is aware} of current
    (pending) interests for particular content, e.g., because it controls
    some NDN routers. Compromised routers that
    receive interests for that content simply inject (satisfy
    interests with) poisoned content, which may then be cached by
    other intervening routers. 
\item Suppose that the adversary {\em anticipates} interests in
    particular content, e.g., a major news-story is about to break on
    CNN or a patch for a popular operating system is about 
    to being released. We also assume that the name of the corresponding content is
    predictable.  The adversary, via numerous distributed
    zombies, issues many near-simultaneous ``legitimate''
    interests for that content. Next, a compromised host or router (that receives
    one or more such interests) replies with 
    poisoned content. Then, caches of routers (that processed preceding interests) 
    become populated with copies of poisoned content.
    Subsequent interests for the same content will return a cached
    version of the same poisoned content.
\end{compactenum}
In~\cite{enhancing-cache-robustness} Xie et al.~consider
a different technique for introducing poisoned content in caches. An 
adversary, who controls a set of zombies, forces them to request 
content produced by the adversary. Such content will take space in 
caches that could otherwise be used more effectively to store ``real'' 
popular content, i.e., this is a locality-disruption attack. 
In this work, we do not consider such attacks, for two reasons: (1)  content 
injected by the adversary is never delivered to consumers who do not 
explicitly request it; and (2) this attack can  be considered as 
legitimate use of NDN; caching policies should be 
designed to deal with this consumer behavior.

While the two aforementioned poisoned content 
attack variants require different adversarial capabilities, 
their impact on the network is almost identical. 
For this reason, we design countermeasures that address the {\em effect} of 
both. 

\subsection{Tentative Countermeasures}
We now discuss tentative countermeasures to content poisoning attacks. 
First we focus on the construction of a strong binding between 
interests and corresponding data packets. We introduce two 
constructions, based on standard NDN features, and analyze their benefits and drawbacks.
Then, we propose further countermeasures based on 
heuristics, inter-router communication and user feedback.

\subsubsection{Self-Certifying and Human-Readable Naming} 
Self-certifying naming~\cite{selfcert} (SCN) allows parties to verify the 
association between a name and the corresponding object without relying on
auxiliary information, such as Public Key Certificates and a PKI.
This makes SCN an effective countermeasure against content poisoning 
attacks~\cite{shenker}.

There are a few well-known approaches in the literature for implementing SCN. The two 
most popular ones are geared for static~\cite{sfs} and dynamic 
content~\cite{selfcert}, respectively. In the former, an object name is computed 
as the hash of its content. In the latter, an object name is constructed as: 
H$(pk)\!:\!L$ where H$(pk)$ is the hash of the producer's public key 
$pk$ and $L$ is a human-readable label.

Users are not expected to handle self-certifying names directly. Instead, SCN 
requires a secure indirection mechanism to map from names familiar to users 
to the corresponding self-certifying names.

NDN uses hierarchical Human-Readable Naming (HRN) 
for content. Human-readable names are designed to be user-friendly, i.e., 
allow consumers to anticipate, guess and remember the name of content they 
wish to retrieve. As discussed in~\cite{VanSm09-Securing}, HRN's advantages over SCN 
can be summarized as:
\begin{compactenum}
\item More efficient routing: SCN provides a flat, location-free namespace, 
which makes it difficult to efficiently retrieve a nearby (cached) copy of 
content corresponding to a particular name~\cite{routing-flat}. Whereas, 
SCN-based architectures resolve names using a location-
independent mechanism, such as DHTs~\cite{stoica-p2p,zhao-faulttolerant};
\item Better usability: consumers can easily 
understand the relationship between an object and its human-readable name;
\item Less complex infrastructure: HRN does not require  the use of a trusted 
name resolution mechanism to map human-readable to network-intelligible names.
\end{compactenum} 
Unfortunately, human readability precludes a strong (i.e., cryptographic) 
binding between a name and a corresponding object. In order to determine 
whether a human-readable name is appropriate for an object, additional 
mechanisms (e.g., a PKI) must be in place.

We consider whether it is possible to integrate the functionalities of SCN 
into NDN, without changing its naming structure. To this end, we introduce 
``Self-Certifying Interests/Data packets'' (SCID), a mechanism that allows 
routers to efficiently and securely determine whether a piece of content is the 
``correct answer'' for particular interest. Two variants of SCID: one for {\em static} 
(S-SCID) and one for {\em dynamic} (D-SCID) content, are described below.

\subsubsection{Static Content}
One of the components automatically appended to the name of each 
NDN data packet upon its creation is a cryptographic hash computed over
its data, name (up to the hash itself) and the signature. A consumer requesting a 
data packet by name, can elect to use this last hash component in an NDN interest.
(Assuming, of course, that the consumer somehow knows this hash ahead of time.)

NDN routers can easily and efficiently determine whether 
a returned data packet corresponds to its requested name with very low overhead.
In fact, routers in the current NDN prototype always verify content hashes.
Our results show that a software implementation of SHA-256 can achieve throughput of
1.5Gbps of traffic (assuming $1,500$-byte packets) on the Intel Core 2 Duo platform, 
as in Section \ref{sec:poisoning}. This is stark contrast with the measly 150Mbps throughput 
we observed in verifying RSA-1024 signatures.

This technique, which we refer to as S-SCID, prevents the adversary from serving
corrupted or fake data packets in response to an interest: the hash of the wrong 
content can not match the one expressed by the consumer.

Linking multiple data packets is quite simple. For example, let 
$CO_1,\ldots,CO_m$ be the collection of data packets corresponding to a 
large file. $CO_i$ includes (in its payload) the hash of $CO_{i+1}$. If
the hash of $CO_1$ can be obtained beforehand, all $CO_i$-s can
be retrieved securely with no danger of fetching the wrong or poisoned content.
The problem is thus reduced to discovering the hash of the initial fragment $CO_1$. 

Also, when fetching a large file, a consumer might wish to have several simultaneously 
outstanding interests, in order to maximize bandwidth usage. Therefore, it is insufficient
for data packets to be singly-linked, as described above. Instead, $CO_i$ 
needs to reference $CO_{i+1},...CO_{i+u}$ where $u$ is the highest number of 
concurrently pending interests. We expect that $u$ is set by the content 
producer based on the nature of specific content. Determining appropriate 
values for $u$ is outside the scope of this paper.

While simple and efficient, S-SCID has several limitations. 
Clearly, a consumer can not be expected to anticipate, guess, remember or recognize the 
hash of content it is about to request. This translates into a 
classical {\em chicken-and-egg} problem. 
The usual SCN solution is to rely on a {\em trusted} 
infrastructure for mapping human-readable to self-certifying names, akin to 
what DNS does today. We discuss how to address this issue (without requiring 
such infrastructures) in the next section.

S-SCID also imposes restrictions on inter-packet dependencies.
In order for packet $A$ to link to packet $B$ ($A\rightarrow B$), the latter must be created and 
named first. This issue makes it impossible for packets to be linked in a cycle, e.g., 
$A\rightarrow B\rightarrow C\rightarrow A$. Consequently, it is unclear how to support
current Web applications that often involve loops in content linkage.
Also, this technique is unsuitable for dynamic content. 
In other words, a consumer has no means of
foretelling the hash of an packet that does not exist at the time of request, e.g., 
the desired packet is the result of a Web search.

\subsubsection{Dynamic Content}
Settings that involve cyclically linked and/or dynamic content require a different flavor of SCID. 
NDN interests include the {\tt PublisherPublicKeyDigest} field, as discussed in 
Section~\ref{sec:ndn}. This field contains the (SHA-256) hash of the public key of
the producer of the matching data packet. Thus, a consumer can (optionally) specify the 
public key that it associates with a desired content name. If this field is 
present in an interest, each intervening NDN router must make sure that 
the corresponding data packet references the same public key. 
We call this technique D-SCID.

Unlike C-SCID, data packets that use D-SCID can include arbitrary 
references to other data packets, including cyclic links or links to 
dynamically generated content. Also, once a consumer learns the hash 
of a producer's public key, it can use it to request all content from that 
producer. Therefore, the nature of links between data packets does not limit the 
number of concurrent pending interests that consumers can issue to retrieve a 
piece of content.

D-SCID prevents adversaries from injecting {\em fake} content in response to an interest.
However, {\em corrupted} content can still be returned as long as it 
references the appropriate producer's public key. (This is, again, because 
NDN routers are not mandated to verify content signatures.)
While S-SCID requires producers to explicitly specify inter-packet links, 
D-SCID does not have such requirement.

\bigskip
Both flavors of SCID combine the benefits of self-certifying and human-readable names.
SCID does not mandate any particular trust model. 
Also, S-SCID and D-SCID are not mutually exclusive.

Let $CO_1, \ldots, CO_m$ be a collection of data packets 
corresponding to a large file, created according to S-SCID -- i.e., each $CO_i$ contains a reference to the hash of $CO_{i+1}$. A consumer first retrieves the content producer's public key $pk$ via its preferred public key distribution mechanism. The hash of $pk$ is used to set the {\tt PublisherPublicKeyDigest} field 
of the interest for $CO_1$. Once $CO_1$ is retrieved, 
the consumer extracts the hash of $CO_2$ from $CO_1$ and issues an  
interests for $CO_2$ using this hash as last component. 
Subsequent interests are issued similarly.

SCN-based architectures generally assume the existence of a trusted 
infrastructure that performs mapping between real-world entities and 
corresponding self-certifying names. Under the same assumption, SCID is a very effective countermeasure against content poisoning attacks; in particular, in the case of static content the exposure of the 
network to such attacks is drastically reduced since
SCID prevents distribution of fake content. As far as corrupted content,
only the first in a collection of packets can be corrupted.

Trust in the first packet of a collection can be bootstrapped using a traditional PKI, as shown in the previous example, or with other  mechanisms such as web of trust~\cite{abd97}, SPKI/SDSI~\cite{spki,sdsi}, etc.

We believe that a combination of the two SCID flavors offers a flexible, trust-model independent solution for securing NDN 
against content poisoning.
To the best of our knowledge, no current SCN-based system allows naming 
content using both ``static'' and ``dynamic'' self-certifying names.

\subsection{Traffic Sampling for Signature Verification}
We now discuss some probabilistic and collaborative techniques
for verifying content signatures by NDN routers.

\descr{Probabilistic Independent Verification.}
Routers verify a random subset of cached content. 
Corrupted packets are immediately removed, while 
those with valid signatures are marked as such 
and never verified again.

Let $r_1,\ldots,r_n$ be a collection of routers. Let $pkt$ be a 
data packet stored in all these routers' caches and $1/v_i$  -- 
the fraction of packets in $r_i$'s cache that are verified at any given time.
$pkt$ is checked by at least one router with probability 
$\mathbb{P} = 1-\prod_{i=1}^n(1-1/v_i)$.

\descr{Probabilistic Disjoint Verification.}
A more effective  strategy involves evenly distributing the 
verification load among a set of routers belonging to the same 
organization. Let $r_1,\ldots,r_n$ be routers in the same organization, 
and let $h_{CO}$ be the least significant 32 bits of the hash of data packet $CO$. 
Router $r_i$ verifies $CO$ if $h_{CO} \equiv i \bmod n $.
Assuming that $h_{CO}$ is distributed uniformly between $0$ and $2^{32}-1$, 
all routers need to verify roughly the same number of packets.

Unfortunately, the adversary can significantly reduce the effectiveness of 
this strategy by generating data packets that are only verified by one router. 
Specifically,  the adversary picks an arbitrary value $x\in[1,n]$, 
creates random data packets and injects them into the network only 
if $h\bmod n =x$.

To prevent this attack, we replace the hash function used to generate $h$ with a 
keyed hash function (HMAC~\cite{hmac}), as follows. All routers belonging to the same 
organization share a secret key $k$. Let $h^k_{CO}$ be the 32 least significant bits of 
HMAC$_k($H$(CO))$. Router $r_i$ verifies $CO$ if $h^k_{CO} \bmod n = i$.
Since HMAC$_k(\cdot)$ is a pseudorandom function, the adversary can 
mount the attack only if it knows the secret key $k$.

Let $1/v_i$ (with $v_i<n$ for all $i\in[1,n]$) be the fraction of cached 
data packets that a router can verify. Given a packet $pkt$, stored in all 
caches of routers in the same organization, $pkt$ is verified with probability 
$\mathbb{P} = 1-\prod_{i=1}^n(1-n/v_i)$.

\descr{Neighbor Verification Feedback.}
To maximize utility of individual router's signature verification, we consider a 
cooperative approach whereby nodes actively exchange information about 
validity of individual data packets.  By having a large number of routers 
verifying packets and cooperating, cryptographic operations can be applied 
less frequently without lowering network's resistance to content poisoning attacks.

Basically, each router (as above) verifies its cached packets probabilistically and
independently.  However, if it determines that a given data packet is corrupted, a 
router issues a special {\em warning} interest on all its interfaces. A warning references
\ndnname{/ndn/warning/hCO}, where \ndnname{/ndn/warning/} corresponds to a
special reserved namespace and \ndnname{hCO} represents the hash of the 
corrupted data packet. The {\tt scope} field of a warning interest is set to 2, 
i.e., this interest type is not forwarded past one hop.

When a router receives a warning interest, it checks whether its cache contains a
referenced packet with hash \ndnname{hCO}. If not, the router discards the warning.  
Otherwise it verifies the content it with some probability $p$ that might depend, 
on its current router CPU load. If signature verification fails, the router issues its own
warning to its neighbors. Otherwise, further warnings from the same interface are 
ignored for a pre-defined period.
To prevent the adversary from injecting fraudulent warnings, every pair of 
adjacent routers could share a symmetric key and use it to authenticate warnings, 
e.g., using a MAC. 

\descr{Consumer Feedback.}
Recall that consumers verify all signatures on data packets.
We take advantage of this property to design a feedback-based verification 
strategy for routers. Consumer feedback can be implemented similarly 
to Neighbor Verification Feedback discussed above, i.e., through specially 
scoped interests. However, allowing consumers to provide feedback 
prompts several new challenges: (1) there is no pairwise trust relationship 
between a router and consumers, even if they are one hop away; (2) consumers
are more likely to be compromised than routers; (3) consumers have almost no 
accountability: it might not be possible to determine which consumer issued a 
false warning.

The intuition behind our strategy is that consumer feedback should not trigger 
immediate action by a router. However, a router should monitor collective (aggregated)
consumer feedback and act whenever its volume exceeds some threshold.

Our strategy is based on a probabilistic trust value $T\in~[0,1]$, associated with 
each content in a router's cache. $T=1$ indicates that the corresponding content 
packet has been verified, while $T\approx0$ indicates that it 
should be selected for verification with probability proportional to $1-T$, 
or deleted if the cache becomes full. New data packets are assigned $T=0.5$. 
This value increases every time the data packet is retrieved, and decreases 
whenever the router receives negative feedback from a consumer.

\section{Related Work \label{sec:relw}}
NDN is an instantiation of the Content-Centric Networking (CCN) paradigm. 
(An alternative term ``Information-Centric Networking'' is largely synonymous.)
Other related architectures include the Data-Oriented Network Architecture
(DONA)~\cite{dona} and TRIAD. 
DONA is based on ``flat'' self-certifying names,
computed as the cryptographic hash of the producer's public key and
a (possibly) human-readable label. Such label, however, is not cryptographically bound
to the content.
New content is published -- i.e., registered -- with a tree of
trusted resolution handlers to  enable retrieval. Resolution handlers
maintain a forwarding table that provides next-hop information for
pieces of content in the network. As such, DONA does not support
dynamically generated content.

Similar to NDN, TRIAD~\cite{triad} names content using
human-readable, location-independent names.
It maps names to available replicas of data using an integrated
directory. It then forwards requests until a copy of the data is found. The
data location is returned to the client, who retrieves it using standard
HTTP/TCP. TRIAD relies on trusted directories to authenticate content
lookups (but not content itself). For additional security, the authors
of~\cite{triad} recommend to limit the network to mutually trusting content
routers.

NDN caching performance optimization has been recently investigated 
with respect  to various metrics including energy
impact~\cite{VanSmBriPlStThBra09-Voice,approximate,greening}.
To the best of our knowledge, the work of Xie, et
al.~\cite{enhancing-cache-robustness} is the  first to address cache
robustness in NDN. It introduces CacheShield, a
mechanism that helps routers to prevent caching unpopular content and
therefore maximizing the use of cache for popular content.

There is lots of previous work on DoS attacks on the current Internet  
infrastructure. Current literature addresses both attacks
and countermeasures on the routing infrastructure~\cite{routing-dos},
packet flooding~\cite{packet-flooding}, reflection
attacks~\cite{reflection-attacks}, DNS cache
poisoning~\cite{And05perilsof} and SYN flooding
attacks~\cite{Wang02detectingsyn}. Proposed 
counter-measures are based on various
strategies and heuristics, including: anomaly
detection~\cite{Carl:2006:DAT:1110639.1110699}, ingress/egress
filtering~\cite{Tupakula:2003:PMC:783106.783137}, IP trace
back~\cite{Lu:2008:GMP:1368310.1368337,Stone:2000:CIO:1251306.1251321},
ISP collaborative defenses \cite{DBLP:journals/tpds/ChenHK07} and user-collaborative defenses~\cite{GkantsidisR06}.

\section{Summary and Future Work\label{sec:conclusions}}
In this paper, we perform initial analysis of NDN's resilience to DoS 
attacks. In doing so, we start by considering attacks on the 
current Internet and assess their impact on NDN. .
We then identify two new type of attacks specific to NDN: interest 
flooding and cache/content poisoning. For type, we discuss effects and 
potential countermeasures. 

Clearly, this paper represents only the first step towards mitigation of
DoS in the context of NDN. Much more work is needed to evaluate the 
effectiveness of proposed countermeasures. In particular, extensive simulation-
and testbed-based experiments must be conducted in order to determine 
optimal parameters for the instantiations of our countermeasures.
Finally, we intend to assess how other content-centric architectures
fare with respect to DoS attacks.

\newpage
\bibliographystyle{plain}
\bibliography{references}

\begin{thebibliography}{10}

\bibitem{sdsi}
Martin Abadi.
\newblock {On {SDSI}'s Linked Local Name Spaces}.
\newblock {\em Journal of Computer Security}, 6(1-2):3--21, October 1998.

\bibitem{abd97}
A~Abdul-Rahman.
\newblock {The PGP Trust Model}, 1997.

\bibitem{hijacking}
Hitesh Ballani, Paul Francis, and Xinyang Zhang.
\newblock {A Study of Prefix Hijacking and Interception in the Internet}.
\newblock {\em SIGCOMM Comput. Commun. Rev.}, 37(4):265--276, August 2007.

\bibitem{hmac}
M.~Bellare, R.~Canetti, and H.~Krawczyk.
\newblock {Keying Hash Functions for Message Authentication}.
\newblock In {\em CRYPTO}, 1996.

\bibitem{routing-flat}
Matthew Caesar, Tyson Condie, Jayanthkumar Kannan, Karthik Lakshminarayanan,
  Ion Stoica, and Scott Shenker.
\newblock Rofl: routing on flat labels.
\newblock In {\em In SIGCOMM}, pages 363--374, 2006.

\bibitem{Carl:2006:DAT:1110639.1110699}
Glenn Carl, George Kesidis, Richard~R. Brooks, and Suresh Rai.
\newblock Denial-of-service attack-detection techniques.
\newblock {\em IEEE Internet Computing}, 10(1):82--89, jan 2006.

\bibitem{content-objects}
{CCNx Content Object}.
\newblock
  \url{http://www.ccnx.org/releases/latest/doc/technical/ContentObject.html}.

\bibitem{interests}
{CCNx Interests Message}.
\newblock
  \url{http://www.ccnx.org/releases/latest/doc/technical/InterestMessage.html}.

\bibitem{DBLP:journals/tpds/ChenHK07}
Yu~Chen, Kai Hwang, and Wei-Shinn Ku.
\newblock Collaborative detection of ddos attacks over multiple network
  domains.
\newblock {\em IEEE Trans. Parallel Distrib. Syst.}, 18(12):1649--1662, 2007.

\bibitem{triad}
D.R. Cheriton and M.~Gritter.
\newblock Triad: A new next-generation internet architecture, 2000.

\bibitem{dns-poisoning}
David Dagon, Manos Antonakakis, Kevin Day, Xiapu Luo, Christopher~P. Lee, and
  Wenke Lee.
\newblock Recursive dns architectures and vulnerability implications.
\newblock In {\em In Network and Distributed System Security Symposium
  (NDSS09)}, 2009.

\bibitem{andana}
S.~DiBenedetto, P.~Gasti, G.~Tsudik, and E.~Uzun.
\newblock {ANDaNA}: Anonymous named data networking application.
\newblock In {\em NDSS}, 2012.

\bibitem{dnssec}
{The DNSSEC Protocol}.
\newblock \url{http://tools.ietf.org/html/rfc2535}.

\bibitem{spki}
Carl~M. Ellison, Bill Frantz, Butler Lampson, Ron Rivest, Brian~M. Thomas, and
  Tatu Ylonen.
\newblock {\em {SPKI} Certificate Theory}, September 1999.
\newblock RFC2693.

\bibitem{selfcert}
David~Mazi Eres, Michael Kaminsky, M.~Frans Kaashoek, and Emmett Witchel.
\newblock Separating key management from file system security.
\newblock In {\em In Proc. SOSP}, pages 124--139, 1999.

\bibitem{sfs}
Kevin Fu, M.~Frans Kaashoek, and David Mazieres.
\newblock Fast and secure distributed read-only file system.
\newblock In {\em ACM Transactions on Computer Systems}, pages 181--196, 2000.

\bibitem{shenker}
Ali Ghodsi, Teemu Koponen, Jarno Rajahalme, Pasi Sarolahti, and Scott Shenker.
\newblock Naming in content-oriented architectures.
\newblock In {\em Proceedings of the ACM SIGCOMM workshop on
  Information-centric networking}, ICN '11, pages 1--6, New York, NY, USA,
  2011. ACM.

\bibitem{GkantsidisR06}
C.~Gkantsidis and P.~Rodriguez.
\newblock Cooperative security for network coding file distribution.
\newblock In {\em INFOCOM 2006. 25th IEEE International Conference on Computer
  Communications, Joint Conference of the IEEE Computer and Communications
  Societies}, 2006.

\bibitem{gritter2001architecture}
M.~Gritter and D.R. Cheriton.
\newblock An architecture for content routing support in the internet.
\newblock In {\em USENIX Symposium on Internet Technologies and Systems}.
  USENIX Association, 2001.

\bibitem{routing-dos}
John Ioannidis and Steven~M. Bellovin.
\newblock Implementing pushback: Router-based defense against ddos attacks.
\newblock In {\em In Proceedings of Network and Distributed System Security
  Symposium}, 2002.

\bibitem{VanSmBriPlStThBra09-Voice}
V.~Jacobson, D.~Smetters, N.~Briggs, M.~Plass, J.~Thornton, and R.~Braynard.
\newblock Voccn: Voice-over content centric networks.
\newblock 2009.

\bibitem{Jacobson2009}
V.~Jacobson, D.~Smetters, J.~Thornton, M.~Plass, N.~Briggs, and R.~Braynard.
\newblock Networking named content.
\newblock {\em The 5th international conference on Emerging networking
  experiments and technologies}, 2009.

\bibitem{packet-flooding}
Jae-Hyun Jun, Hyunju Oh, and Sung-Ho Kim.
\newblock Ddos flooding attack detection through a step-by-step investigation.
\newblock {\em Networked Embedded Systems for Enterprise Applications, IEEE
  International Conference on}, 0:1--5, 2011.

\bibitem{koponen2007data}
T.~Koponen, M.~Chawla, B.G. Chun, A.~Ermolinskiy, K.H. Kim, S.~Shenker, and
  I.~Stoica.
\newblock A data-oriented (and beyond) network architecture.
\newblock {\em ACM SIGCOMM Computer Communication Review}, 37(4):181--192,
  2007.

\bibitem{dona}
Teemu Koponen, Mohit Chawla, Byung-Gon Chun, Andrey Ermolinskiy, Kye~Hyun Kim,
  Scott Shenker, and Ion Stoica.
\newblock A data-oriented (and beyond) network architecture.
\newblock In {\em SIGCOMM '07}, pages 181--192, New York, NY, USA, 2007. ACM.

\bibitem{lad-PHAS}
Mohit Lad, Dan Massey, Dan Pei, Yiguo Wu, Beichuan Zhang, and Lixia Zhang.
\newblock Phas: a prefix hijack alert system.
\newblock {\em USENIX Security}, August 2006.

\bibitem{greening}
Uichin Lee, Ivica Rimac, and Volker Hilt.
\newblock Greening the internet with content-centric networking.
\newblock In {\em e-Energy}, pages 179--182, 2010.

\bibitem{Lu:2008:GMP:1368310.1368337}
Liming Lu, Mun~Choon Chan, and Ee-Chien Chang.
\newblock A general model of probabilistic packet marking for ip traceback.
\newblock In {\em Proceedings of the 2008 ACM symposium on Information,
  computer and communications security}, ASIACCS '08, pages 179--188, New York,
  NY, USA, 2008. ACM.

\bibitem{NDN}
Named data networking project {(NDN)}.
\newblock \url{http://named-data.org}.

\bibitem{reflection-attacks}
Vern Paxson.
\newblock An analysis of using reflectors for distributed denial-of-service
  attacks.
\newblock {\em SIGCOMM Comput. Commun. Rev.}, 31(3):38--47, July 2001.

\bibitem{And05perilsof}
V.~Ramasubramanian and E.~Sirer.
\newblock Perils of transitive trust in the domain name system.
\newblock In {\em Proc. International Measurement Conference}, 2005.

\bibitem{approximate}
Elisha~J. Rosensweig, Jim Kurose, and Don Towsley.
\newblock Approximate models for general cache networks.
\newblock In {\em Proceedings of the 29th conference on Information
  communications}, INFOCOM'10, pages 1100--1108, Piscataway, NJ, USA, 2010.
  IEEE Press.

\bibitem{VanSm09-Securing}
Diana~K. Smetters and Van Jacobson.
\newblock Securing network content.
\newblock Palo Alto Research Center, 2009.

\bibitem{stoica-p2p}
Ion Stoica, Robert Morris, David Karger, M.~Frans Kaashoek, and Hari
  Balakrishnan.
\newblock Chord: A scalable peer-to-peer lookup service for internet
  applications.
\newblock pages 149--160, 2001.

\bibitem{Stone:2000:CIO:1251306.1251321}
Robert Stone.
\newblock Centertrack: an ip overlay network for tracking dos floods.
\newblock In {\em Proceedings of the 9th conference on USENIX Security
  Symposium - Volume 9}, SSYM'00, pages 15--15, Berkeley, CA, USA, 2000. USENIX
  Association.

\bibitem{Tupakula:2003:PMC:783106.783137}
Udaya~Kiran Tupakula and Vijay Varadharajan.
\newblock A practical method to counteract denial of service attacks.
\newblock In {\em Proceedings of the 26th Australasian computer science
  conference - Volume 16}, ACSC '03, pages 275--284, Darlinghurst, Australia,
  Australia, 2003. Australian Computer Society, Inc.

\bibitem{Wang02detectingsyn}
Haining Wang, Danlu Zhang, and Kang~G. Shin.
\newblock Detecting syn flooding attacks.
\newblock In {\em In Proceedings of the IEEE Infocom}, pages 1530--1539. IEEE,
  2002.

\bibitem{enhancing-cache-robustness}
M.~Xie, I.~Widjaja, and H.~Wang.
\newblock Enhancing cache robustness for content-centric networks.
\newblock In {\em In Proceedings of the IEEE Infocom}, 2012.

\bibitem{zhao-faulttolerant}
Ben~Y. Zhao, John~D. Kubiatowicz, and Anthony~D. Joseph.
\newblock Tapestry: An infrastructure for fault-tolerant wide-area location and
  routing.
\newblock Technical Report UCB/CSD-01-1141, EECS Department, University of
  California, Berkeley, Apr 2001.

\end{thebibliography}

\end{document}